# On strongly coupled quenched QED4, again: chiral symmetry breaking, Goldstone mechanism and the nature of the continuum limit * †

M–P Lombardo [a] A. Kocić [b] and J. B. Kogut[b]

[a]HLRZ c/o KFA Jülich, D-52425 Jülich, Germany

[b]Department of Physics, University of Illinois at Urbana–Champaign, 1110 West Green Street, Urbana, IL 61801-3080, U.S.A.

We explore the possibility of a trivial continuum limit of strongly coupled quenched QED4 by contrasting our results with a Nambu–Jona Lasinio equation of state. The data does not compare favorably with such scenario. We study in detail the interplay of chiral symmetry breaking with the Goldstone mechanism, and clarify some puzzling features of past results.

## 1. INTRODUCTION AND SUMMARY

We present a comprehensive analysis of old and new results for the order parameter, the susceptibilities and the spectrum of lattice QED4 in the quenched approximation [1]. The nature of the continuum limit – interacting vs. trival – is studied by contrasting the results for $<\bar{\psi}\psi>$ and its susceptibilities with the predictions of two critical scenarios –powerlaw scaling, and logarithmically corrected mean–field. The hypothesis of a power law scaling has already been tested in past work, where scaling laws characteristic of an interacting, non–trivial underlying theory were found. However, the theoretically motivated triviality à la Nambu–Jona Lasinio (as opposed to $\lambda\phi^4$ [2]) has not been explored in detail, after preliminary analysis of [3], where this proposal was first put forward. Thus, the consistency of the results with a logarithmically trivial equation of state was still an open problem. One of the purposes of this work is to discriminate among power–law scaling and triviality realized à la NJL. We have found that a NJL motivated EOS does not compare favourably with the data. A byproduct of our analysis, which used new techniques, is a very accurate check of previous results for the critical exponents. As extensively discussed in the past, the critical exponents are consistent with a particular point on the critical line of the continuum model [4]. Then, to further sharpen our understanding of the lattice model we need to learn how to move along the critical line. The contributions of S. Hands [5] and J.-F. Lagaë [6] to these Proceedings address this issue.

We also aimed to clarify the interplay of chiral symmetry breaking with the Goldstone mechanism. The pion mass, measured in the past, displayed some rather puzzling systematics: it decreased with $\beta$ even as we pass through the chiral restoration transition. Here, we study the behaviour of scalar and pseudoscalar propagators in much more detail than in past work. The numerical results for the pion mass are confirmed. Nevertheless, we observe clear indications of chiral symmetry restoration, both in the scalar and in the vector sector: the propagators of the chiral doublets become degenerate, and the amplitude of the Goldstone mode drops by an order of magnitude across the transition. We speculate that in the thermodynamic limit, the Goldstone mode would decouple completely at the chiral symmetry restoration transition. We have thus gained a better understanding of the interplay of chiral symmetry breaking and the Goldstone mechanism, and clarified some puzzling features of past results. The major open problem in the spectroscopic sector remains the $\sigma$–fermion level ordering.

*Partially supported by NSF under grant NSF-PHY92-00148. Simulations done on the C90 CRAY's at PSC and NERSC, and on the CM-2 at PSC
†Contribution to Lattice94, Bielefeld

## 2. EQUATION OF STATE AND SCALING

We have considered the power-law equation of state:

$$m_e = a^P t <\bar\psi\psi> + b^P <\bar\psi\psi>^\delta \quad (1)$$

in parallel with the NJL log-corrected mean field equation of state:

$$m_e = a^L t <\bar\psi\psi> + b^L <\bar\psi\psi>^3 \, log <\bar\psi\psi> \quad (2)$$

$(t = (\beta_c - \beta))$

For the direct fits to the EOS's, we have used our old data for the chiral condensate obtained on the $24^4$ lattice. We have fitted the data for several $\beta$ ranges. The power-law fits are satisfactory and stable with respect to the fitted interval once $\beta$ is restricted to the range .240 to .260. The fit à la Nambu–Jona-Lasinio, on the other hand, are qualitatively poorer, and the resulting parameters are sensitive to the $\beta$ interval chosen. The fits, whose results are summarized in Table 1, clearly favour power-law behavior, with $\beta_c = .2573(1)$, $\delta = 2.13(1)$, in agreement with previous findings.

The results for the critical exponents can be obtained very nicely in graphic form, under the *single* assumption of the existence of an Equation of State which obeys scaling: Consider the two logarithmic derivatives $R_t = \frac{t}{<\bar\psi\psi>} \frac{\partial <\bar\psi\psi>}{\partial t}$ and $R_m = \frac{m}{<\bar\psi\psi>} \frac{\partial <\bar\psi\psi>}{\partial m}$. $R_t$ and $R_m$ satisfy:

$$\frac{R_t}{\beta_{mag}} + \delta R_m = 1 \quad (3)$$

In ref. [7] we have discussed the properties of the ratio $R_m(t,m)$. It was shown how its behaviour is dictated by symmetry arguments, and provides information on the critical point, and critical exponents, with no a priori assumptions. In particular, $R_m(t,0) = 0(1)$ in the strong (weak) coupling limit and $R_m(0,m) = 1/\delta$. Thus, it is possible to read off the exponents from the plot $R_t$ vs. $R_m$ (Fig.1). We have also obtained direct estimates of $R_m$ via a stochastic estimator, on $8^4, 16^4, 16^3 \times 32$ lattices, for a wide array of masses and couplings. Finite size and geometry have been carefully monitored, and we concluded that the data on the $16^4$ lattices are

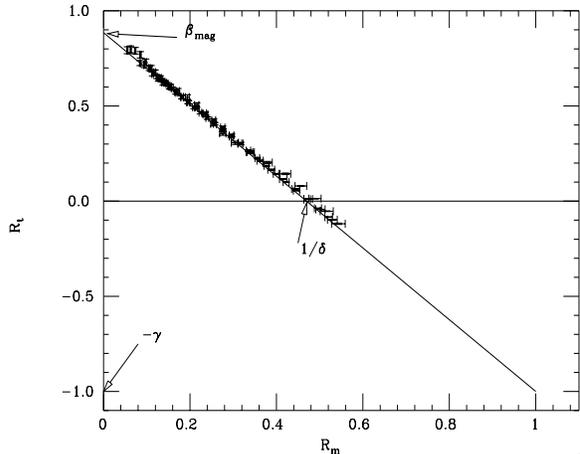

Figure 1. $R_t$ vs $R_m$. The results are from the $24^4$ lattice, the derivatives are evaluated numerically. The input value for $\beta_c$ was .2572. The straight line is the prediction of the power-law EOS

systematic–free. We show in Fig. 2 the ratio of susceptibilities plotted at fixed $\beta$ as a function of the bare electron mass. We plot our results on the $16^4$ lattice, and, as a cross check, the results on the $24^4$ lattice, where the sigma susceptibility was obtained by numerical differentiation, wherever they overlap. While for power law scaling $R_m$ is expected to be independent of $m$ and equal to $1/\delta$ at the critical point, for the NJL model it would follow the 'effective' $\delta$, $R_m(0,m) = 1/\delta_{eff} = 1/(3 + 1/log <\bar\psi\psi>)$. The straight line is drawn giving $1/\delta$ as estimated by the fit, and falls, as it should, half way between $\beta = .255$ and $\beta = .260$.

## 3. SPECTROSCOPY

We have obtained results on a $16^3 \times 32$ lattice, at $\beta = (.250\ .255\ .260\ .270\ .280)$, for bare quark masses ranging from .003 to .025. The analysis of the scalar and pseudoscalar propagators provides information on the mechanism of chiral symmetry breaking and the appearance of a Goldstone mode in the meson spectrum. Consider for instance the amplitude in the fundamental channel, as obtained from a two-particles fit of the pseudoscalar propagator, and the total amplitude minus the fundamental one. We show in Fig. 3 the results for $m = .003$ and $m = .009$. The change in the trend around the transition is very clear, the effect being more pronounced at small masses, as it should be. This explains how the low-lying state in the pseudoscalar channel maintains the property of a Goldstone boson beyond the transition point, as noticed in past work, but there is no contradiction with the Goldstone Theorem. The candidate Goldstone boson simply decouples in the weak coupling phase. We have also observed clear signatures of chiral symmetry restoration in the propagators of the chiral doublets, which become (almost) degenerate at the transition.

Table 1
Results of the EOS fits

| $\beta$ range | $a^L$ | $b^L$ | $\beta_c^L$ | $a^P$ | $b^P$ | $\beta_c^P$ | $\delta$ |
|---|---|---|---|---|---|---|---|
| .250–.260 | -5.43(4) | -2.76(4) | .2542(1) | -5.40(2) | 1.04(14) | .2572(6) | 2.12(8) |
| .245–.260 | -5.41(3) | -2.59(3) | .2538(1) | -5.42(2) | 1.05(2) | .2572(1) | 2.13(1) |
| .240–.260 | – | – | – | -5.54(2) | 1.07(2) | .2573(1) | 2.13(1) |

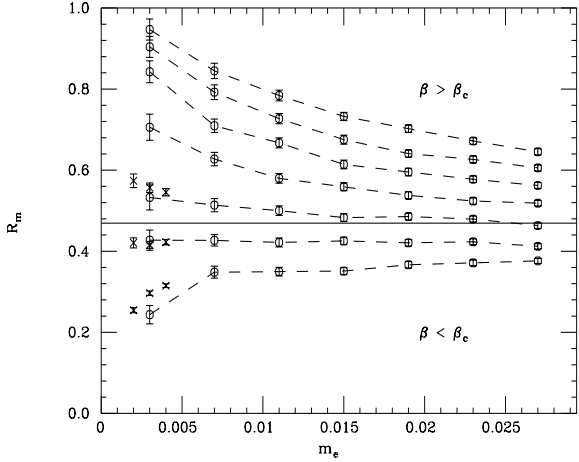

Figure 2. $R_m$, $16^4$ lattice, circles. $\beta = .250$–.280 from bottom to top. The crosses are for $R_m$ evaluated by taking the derivative numerically on the $24^4$ lattice. The straight line is drawn at $1/\delta$, with $\delta = 2.12$, and shows that $.255 < \beta_c < .260$ in agreement with the results of the EOS fit.

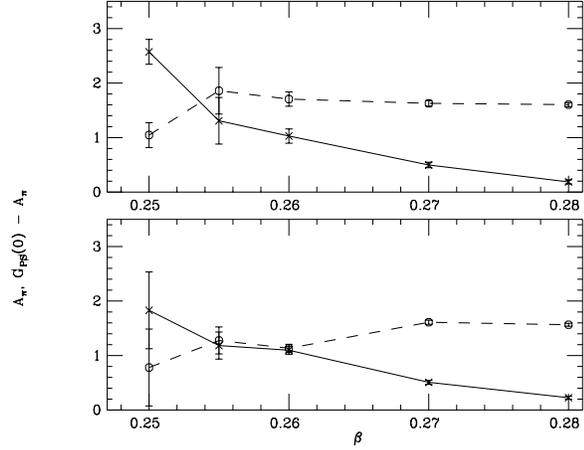

Figure 3. Amplitude of the Goldstone mode (solid), and sum of the amplitudes of the excited states (dash) for $m_e = .003$ (top) and $m_e = .009$ (bottom), as a function of $\beta$. The Goldstone mode becomes dominant around the critical $\beta$, the effect being more pronounced at small mass.

**REFERENCES**

1. M.-P. Lombardo, A. Kocić and J. B. Kogut, *More on strongly coupled quenched QED*, hep-lat/9411051, contains a detailed exposition of these results, and a comprehensive set of references.
2. A. Kocić, Nucl. Phys. **B34** (Proc. Suppl.) (1994) 123
3. A. M. Horowitz, Nucl. Phys. **B17** (Proc. Suppl.)(1990) 694
4. A. Kocić, S. Hands, J. B. Kogut and E. Dagotto, Nucl. Phys. **B347** (1990) 217; W. A. Bardeen, S. Love and V. A. Miransky, Phys. Rev. **D42** (1990) 3514
5. S. Hands, these Proceedings
6. J.-F. Lagaë, these Proceedings
7. A. Kocić, J. B. Kogut and M.-P. Lombardo, Nucl. Phys. **B398** (1993)376